\documentclass{pasa}%
\usepackage{color}
\title[LPVs in Globular Clusters]{A search for long period variables in Globular Clusters: M22 and IC4499}
\author[Sahay, Lebzelter and Wood]{Sahay, A.\\  
  \affil{Department of Applied Physics, Indian School of Mines Dhanbad}
\and Lebzelter, T.\\
  \affil{University of Vienna, Department of Astrophysics, T\"urkenschanzstrasse 17, A1180 Vienna, Austria}
\and Wood, P.R.\\
  \affil{Australian National University, Research School for Astronomy and Astrophysics}}
\jid{PASA}
\doi{10.1017/pas.\the\year.2013}
\jyear{\the\year}

\begin{document}%
\begin{abstract}
We report on the results of a long time photometric monitoring of the two metal poor Galactic globular clusters
M22 and IC4499 searching for long period variables (LPVs) on the upper giant branch. We detected 22 new LPVs
in the field of M22 and confirmed the variability of six known variables. Periods could be determined
for 16 of them. In the field of IC4499 we detected and characterized 2 new LPVs. Cluster membership is evaluated for
all the variables based on photometry and literature data, and the location of the stars in log~P-$K$-diagram 
is discussed. Our findings give further support to the presence of LPVs at metallicities as low as [Fe/H]$=-$1.7. The luminosity range
where LPVs are found in metal poor clusters is lower than in more metal rich clusters.
\end{abstract}
\maketitle%
\section{INTRODUCTION }
\label{sec:intro}
The Asymptotic Giant Branch (AGB) phase is a key stage in the stellar evolution of low and intermediate mass stars.
It is characterized by nucleosynthesis, mixing, and mass loss events, and it is 
therefore of fundamental importance for the
cosmic matter cycle. The atmospheres of these stars are highly extended -- not least due to the levitation of
material by stellar radial pulsation -- and this presents major challenge for determining the parameters of these objects
\cite{LNH10}.  A study of the variability of AGB stars provides another mechanism for determining
the parameters of red giant stars, particularly their mass, their cumulative amount of mass loss
and possibly gross abundance variations such as a high helium content.

An interesting possibility is to study AGB stars 
in single stellar populations where metallicity, mass (age), and distance
can be determined from other objects in that system. Therefore, two of us (TL and PW) have started a long time 
photometric monitoring programme of several Galactic and Magellanic Clouds globular clusters to get 
an extensive
repository of AGB stars in these systems and to use the detected sample in each cluster to further investigate the
properties of low and intermediate mass stars during this final evolutionary stage. The results on several of the
clusters studied have been published already: NGC 104 \cite{LW05}, NGC 1846 \cite{LW07}, NGC 1978 and NGC 419
\cite{KWSL10}, and NGC 362 and NGC 2808 \cite{LW11}. Here we present results on two more clusters, 
namely M22 (NGC 6656) and IC 4499.

M22 (l=9.9 deg, b=$-$7.6 deg) is a galactic globular cluster at a distance of $\sim$3.1 kpc from the sun 
and a mean metallicity of [Fe/H]$=-$1.64 \cite{H96}. An age estimate of 12.5 Gyr \cite{vdB13} places it
among the bulk of the Galactic globular clusters. An average reddening value of $E(B-V)=$0.38 has been
found with significant variations across the cluster \cite{Kud13}.
It was noticed quite early \cite{H77} that the colour-magnitude diagram of this cluster shows an unusual colour
spread of the giant branch. This triggered a large number of spectroscopic investigations of the cluster giants which indeed
found indications for a metallicity spread similar to the case of $\omega$~Cen. Early studies measuring metallicities in
a few giants only did not come to conclusive results \cite{Cohen81,Pila82}. The investigation by Da Costa et al. \shortcite{da09}
on 55 red giants showed a broad range of metallicities between [Fe/H]$=-$1.9 and [Fe/H]$=-$1.45, in agreement with earlier
papers by Lehnert et al.~\shortcite{L91} or Norris and Freeman \shortcite{NF83}.  The metallicity spread 
was later confirmed by Alves-Brito et al.~\shortcite{Alves12}. The works of
Marino and collaborators \cite{M09,M11} gave a strong indication for the existence of two populations with a separation in
[Fe/H] of 0.15 dex. The two groups differ also in their abundances of s-process elements, the more metal rich ones showing
a higher abundance of s-process dominated species. This is in agreement with an evolutionary scenario of the cluster where a second generation
of stars is enriched in neutron capture elements by AGB stars \cite{M11}. The split was also found in a sample of M22 subgiants
\cite{M12}.
While the presence of two populations in M22 or at least a metallicity spread 
is now widely accepted in the literature, we note that there are also a few
papers finding contradicting results, i.e. the existence of only one population \cite{A95,MPFB04,I04}.


Clement et al.~\shortcite{Cl01} list 72\footnote{Including the 2009 update published on their webpage: 
http://www.astro.utoronto.ca/~cclement/cat/C1833m239.} variable stars known to date in the cluster field. Most of them are RR Lyr stars, 
but 12 are probably red 
long-period variables, for some of which periods are given. V5, V8, and V9 are marked as cluster members, while for the others
the membership is not clear.
Recent surveys for variable stars in that cluster include the study by Kaluzny and Thompson \shortcite{KTh01} identifying several SX Phe stars and candidate eclipsing binaries. 
Eight additional variables were identified in the cluster halo by Kravtsov et al.~\cite{K94}.

While M22 belongs to the most massive globular clusters, a further target of our survey, IC4499, is a sparse cluster of rather low density in the outer halo of the Galaxy.
We combine here our results of this cluster with the ones for M22 since both clusters have a similar low mean metallicity. Harris \shortcite{H96}
gives its distance from the sun as 17.6\,kpc, while Piotto et al.~\shortcite{Pio02} estimate a reddening value $E(B-V)=$0.23. 
These values for distance and reddening are in good agreement
with results from RR Lyr variables \cite{Storm04}.
Only a few spectroscopic metallicity determinations are available -- see Hankey and Cole \shortcite{HC11} for a recent
summary. Values for [Fe/H] range between $-$1.5 and $-$1.65 \cite{S82,WN96} on the Zinn \& West \shortcite{ZW84} scale.
There exists also a number of photometric attempts to determine the cluster metallicity which give somewhat lower 
values around [Fe/H]$=-$1.75 \cite{FP95,F95}.
Hankey and Cole themselves present an extensive spectroscopic study of 43 red giants finding
a metallicity of [Fe/H]$=-$1.52$\pm$0.12. 
There is some debate that IC4499 may be 2-4 Gyr younger than the bulk of the metal poor globular clusters
\cite{F95,SW02,WKA11}.
IC 4499 is famous for its high number of RR Lyr stars \cite{S93,Cl01,WN96}.  However,
no long period variables (LPVs) are known in this cluster according to Clement et al.~\shortcite{Cl01}.


\section{DATA ANALYSIS}
The two clusters are part of our observing program to identify and characterize LPVs in Galactic globular 
clusters. The observations of M22 and IC 4499 were obtained and analysed the same way as the previous papers in this series 
(see Section~\ref{sec:intro}). 
Thus we will give here only a brief summary and refer to Lebzelter and Wood \shortcite{LW05} for details.
Our monitoring program of IC4499 and M22 started in May 2002 at the 50 inch telescope at Mount Stromlo. This 
telescope was equipped with a two channel camera used earlier for the MACHO experiment \cite{A93}. 
Observations were obtained once to twice a 
week, but ended unexpectedly after a few months when Mount Stromlo observatory was destroyed by a bush 
fire. Altogether we collected 21 usable frames for M22 and 27 usable frames for IC4499 over this time span. All 
observations were done in queue observing mode.

Data from both the blue and the red channel were used for the detection of variables and the determination of light curves. Due to the 
larger light amplitude of long period variables in the blue than in the red, the blue frames received a higher weight for 
the detection of variables and the determination of periods. Since the two channels had the same exposure time and since
LPVs are typically the brightest stars in the red in an old stellar cluster, several of our variables were
over-exposed on the red frames and thus only the blue channel data were useable. 
Small positional shifts of the cluster on the frame between the observing nights and the presence of an area of
bad pixels on the detectors prevented the measurement of all the stars on all our frames. 

Variable star detection and extraction of the light curve data was done with the help of 
the image subtraction code ISIS 2.1 developed by Alard \shortcite{A00}. 
Stellar fluxes on the reference 
frames were measured using standard IRAF software. Two images observed in the same night give a differential 
photometric accuracy for a bright cluster giant around 0.005 mag in the blue. 
To determine the photometric zero point in $V$, 
we identified several stars with known
$V$ photometry from the survey by Monaco et al.\,\shortcite{MPFB04} on the blue frames. No correction has been applied for the
difference between Johnson $V$ and MACHO blue. Since the comparison stars were of 
colour similar to most of
our target stars, we expect a rather small difference of around 0.1 magnitudes 
\cite{BG99}. To calibrate the red MACHO frames, $R$ band magnitudes would have been required. Since these
were not available in our case, the red frames were analysed with an arbitrary zero point.

Among the variable stars detected, we selected the long period variables based on the brightness (on the upper giant branch), 
timescale of the variation (more than 30 days), and a total light amplitude in V of at 
least 0.03\,mag. As in the previous papers of this series, Period98 \cite{S98} was used to derive periods from our
light curves. Period98 is a code which can 
compute a discrete Fourier 
transformation in combination with a least-squares fitting of multiple frequencies on the data. A maximum of two 
periods was considered for each star.  For the period range studied here (a few ten days) we do not expect
a significant aliasing problem from our sampling of the light curves.  Note that, given that the periods of semiregular
variables may change from time to time, our results only describe 
the light variation at the time of observation. This may represent only a small part of a more
complex light change. In particular, long time variations on time scales of a year or more are not accessible
with our data set.

\begin{table*}
\caption{Long period variabls in the field of M22}
\begin{center}
\begin{tabular*}{\textwidth}{@{}l\x c\x c\x c\x c\x c\x c\x c\x c\x c\x c@{}}
\hline\hline
Name & RA(2000) & Dec(2000) & $V_{mean}$ & $I^a$ & J & H & K & Period & $\Delta$V & Remark\\
~ & ~ & ~ & [mag] & [mag] & [mag] & [mag] & [mag] & [d] & [mag] & ~ \\
\hline%
V5 & 18 36 10.57 & -23 55 00.5 & 11.16 & 9.18 & 7.72 & 6.99 & 6.731 & 54 & 0.8 & P$_{lit}$~93\,d\\
V14 & 18 36 40.52 & -23 46 07.4 & 14.84 & 10.96 & 8.924 & 8.037 & 7.515 & 202$^c$ & 2.2 & V1266 Cyg, P$_{lit}$ 202\,d \\ 
 &               &               &      &      &        &        &    &            &        & likely non-member\\
V30 & 18 36 41.05 & -23 58 19.5 & 11.34 & 9.19 & 7.75 & 6.991 & 6.759 & 62 & 0.2 & P$_{lit}$ 82.5\,d\\
V34$^{b}$ & 18 36 26.07 & -23 55 33.9 & 12.41 & 9.48 & 8.019 & 7.225 & 6.988 & 63: & 0.2: & no P$_{lit}$\\ 
V35 & 18 36 24.04 & -23 54 29.3 & 12.03 & 9.41 & 7.948 & 7.171 & 6.949 & 56: & 1.0 & no P$_{lit}$\\ 
SLW1 & 18 35 41.00 & -23 53 28.8 & 12.34 & 11.06 & 9.836 & 9.092 & 8.896 & 61 & 2.2 & likely non-member \\ 
SLW2 & 18 35 59.09 & -23 51 34.7 & 13.90 & 10.77 & 9.143 & 8.082 & 7.816 & 105 & 1.2 & likely non-member\\ 
SLW3 & 18 35 59.44 & -23 54 02.7 & 15.92 & 12.72 & 11.209 & 10.328 & 10.06 & 77: & 1.0 & likely non-member\\ 
SLW4 & 18 36 17.51 & -23 54 26.3 & 11.41 & 9.32 & 7.82  & 7.011 & 6.783 & 57 & 1.2 & ~ \\ 
SLW5 & 18 36 18.38 & -23 54 01.3 & 11.15 & 9.62 & 8.289 & 7.53 & 7.306 & 76 & 0.03 & ~ \\ 
SLW6 & 18 36 19.27 & -23 53 26.7 & 11.18 & 9.53 & 8.013 & 7.688 & 7.257 & 73: & 0.07 &  \\ 
SLW7 & 18 36 21.01 & -23 54 42.5 & 11.28 & 9.68 & 8.407 & 7.614 & 7.438 & 61 & 0.1 & ~ \\ 
SLW8 & 18 36 21.64 & -23 55 57.0 & 11.90 & 9.36 & 7.933 & 7.163 & 6.929 & 61: & 1.2 & ~ \\ 
SLW9 & 18 36 25.42 & -23 54 35.6 & 13.23 & 11.60 & 10.512 & 9.878 & 9.689 & 122: & 0.6 & ~ \\ 
SLW10 & 18 36 26.64 & -23 45 02.9 & 13.71 & 11.51 & 9.432 & 8.498 & 8.104 & 46$^d$ & 0.03 & ~\\ 
SLW11 & 18 36 28.05 & -23 53 23.2 & 11.39 & 9.55 & 8.189 & 7.34 & 7.197 & 79 & 0.07 & non-member? \\ 
\hline\hline
\end{tabular*}
\end{center}
\tabnote{$^aI$-band data taken from the DENIS database \cite{Denis05}.}
\tabnote{$^b$Classification as LPV doubtful.}
\tabnote{$^c$Literature value used.}
\tabnote{$^d$Long secondary period.}
\label{tab1}
\end{table*}

\begin{table*}
\caption{Long period variable candidates of M22 with unknown periods} 
\begin{center}
\begin{tabular*}{\textwidth}{@{}l\x c\x c\x c\x c\x c\x c\x c\x c@{}}
\hline \hline
Name & RA(2000) & Dec(2000) & $V_{mean}$ & $I^a$ & J & H & K \\
~ & ~ & ~ & [mag] & [mag] & [mag] & [mag] & [mag] \\
\hline 
V9$^b$ & 18 36 08.19 & -23 55 02.8 & 11.06 & 9.34 & 7.701 & 6.926 & 6.676 \\ 
SLW12 & 18 36 02.18 & -23 56 50.6 & 12.35 & 10.57 & 9.427 & 8.755 & 8.583  \\ 
SLW13 & 18 36 10.22 & -23 48 44.4 & 11.94 & 9.35 & 7.842 & 7.005 & 6.737 \\ 
SLW14 & 18 36 14.57 & -23 52 45.4 & 13.40 & 12.36 & 11.37 & 10.794 & 10.708 \\ 
SLW15 & 18 36 20.64 & -23 51 36.0 & 12.89 & 10.01 & 8.822 & 8.072 & 7.868 \\ 
SLW16 & 18 36 23.47 & -23 54 53.8 & 11.23 & 9.40 & 7.988 & 7.23 & 6.971 \\ 
SLW17 & 18 36 26.13 & -23 54 51.0 & 11.32 & 10.04 & 8.752 & 8.013 & 7.881 \\ 
SLW18 & 18 36 26.25 & -23 52 32.0 & 12.79 & 11.26 & 10.161 & 9.491 & 9.352 \\ 
SLW19 & 18 36 26.27 & -23 51 48.3 & 12.62 & 10.82 & 9.782 & 9.136 & 9.011 \\ 
SLW20 & 18 36 30.53 & -23 53 57.8 & 11.59 & 9.92 & 8.697 & 8.007 & 7.827 \\ 
SLW21 & 18 36 33.01 & -23 54 35.6 & 11.28 & 10.09 & 9.114 & 8.589 & 8.401 \\ 
SLW22 & 18 36 33.01 & -23 42 44.1 & 14.15 &    & 8.719 & 7.817 & 7.469 \\ 
\hline \hline
\end{tabular*}\label{tab2}
\end{center}
\tabnote{$^aI$-band data taken from the DENIS database \cite{Denis05}.}
\tabnote{$^b$P$_{lit}$~87.7\,d.}
\end{table*}

\begin{table*}
\caption{Long period variables of IC 4499} 
\begin{center}
\begin{tabular*}{\textwidth}{@{}l\x c\x c\x c\x c\x c\x c\x c\x c\x c@{}}
\hline \hline
Name & RA(2000) & Dec(2000) & $V_{mean}$ & $I^a$ & J & H & K & Period & $\Delta$V\\
~ & ~ & ~ & [mag] & [mag] & [mag] & [mag] & [mag] & [d] & [mag]\\
\hline
SLW1 & 15 00 22.74 & -82 12 35.9 & 12.19 & 11.19 & 10.203 & 9.587 & 9.424 & 67 & 0.8 \\ 
SLW2 & 15 01 20.37 & -82 12 43.5 & 13.44 & 11.71 & 10.805 & 10.183 & 10.027 & 35 & 0.8 \\
\hline \hline
\end{tabular*}\label{tab3}
\end{center}
\tabnote{$^aI$-band data taken from the DENIS database \cite{Denis05}.}
\end{table*}

\section{RESULTS}
\subsection{M22}
 Five known and eleven new LPVs were found in M22.  
All stars are identified in Table \ref{tab1}, where 
coordinates, $V$, DENIS $I$, and 2MASS magnitudes, periods and amplitudes are listed. 
As in Lebzelter \& Wood \shortcite{LW05}
we used the variable star identifiers given by Clement et al.~\shortcite{Cl01} 
where available. All other variables were 
named SLW with some number. Uncertain periods or amplitudes are marked by a colon.
The previously reported LPV candidates
V8,  V17, V26, V28, and V31 to V33 were outside our studied field of view.

Three of the previously known LPVs in Table \ref{tab1} have periods listed in
the literature which is given in the last column of Table\,\ref{tab1}. In the case of V14
-- originally detected by Bailey \shortcite{B02} --
we used the literature period of 202 days, since the length of our time series is not
sufficient to cover one complete light cycle. We can confirm that the literature period is
in a good agreement with our data in this case. For the other two stars with periods from
the literature, however, we find quite different periods. In the case of V5 we note some irregular
behaviour at the beginning of our light curve, but the second half clearly supports a period of
54\,d, i.e.~roughly half the literature value. 
On the other hand, Wehlau and Sawyer-Hogg \shortcite{WS77} found a 93 day
period to be the most prominent one over a very long time interval.
This star appears to show multiple periods.

For V30 the longer period found in the literature seems to be unlikely, although the situation 
is not as clear as for V5. A reason for the differences in period with former studies for these three stars 
is unknown. The period might indeed have changed, but it may also be that the time sampling on which the literature
values are based was not sufficient since in all cases we get shorter periods.
We note that there is a star close to our variable SLW9 that has been reported variable by 
Kaluzny and Thompson~\shortcite{KTh01}. It is not clear if their star pk6 is different from SLW9. However,
Kaluzny and Thompson classified it as an eclipsing binary. 
We find no obvious indications for binarity in our data, therefore we suppose that Kaluzny and Thompson
discovered a different variable.

We further note that the star V9 in this cluster has been previously identified as red variable. Despite
the star having colour and brightness similar to other long period variables in our sample, its variability 
amplitude in $V$ in our data is very small and the star
does not show any characteristics of a long period light change. The period value given in the literature
could not be confirmed by our data, although the Wehlau and Sawyer-Hogg \shortcite{WS77} measurements
seem to support the variability of this star. We suspect that V9 is currently in a phase of quiescence
occasionally found in semiregular variables, and the Wehlau and Sawyer-Hogg data indeed show indications 
for a variable amplitude.

\begin{figure*}
\begin{center}
\includegraphics[width=42pc,height=42pc]{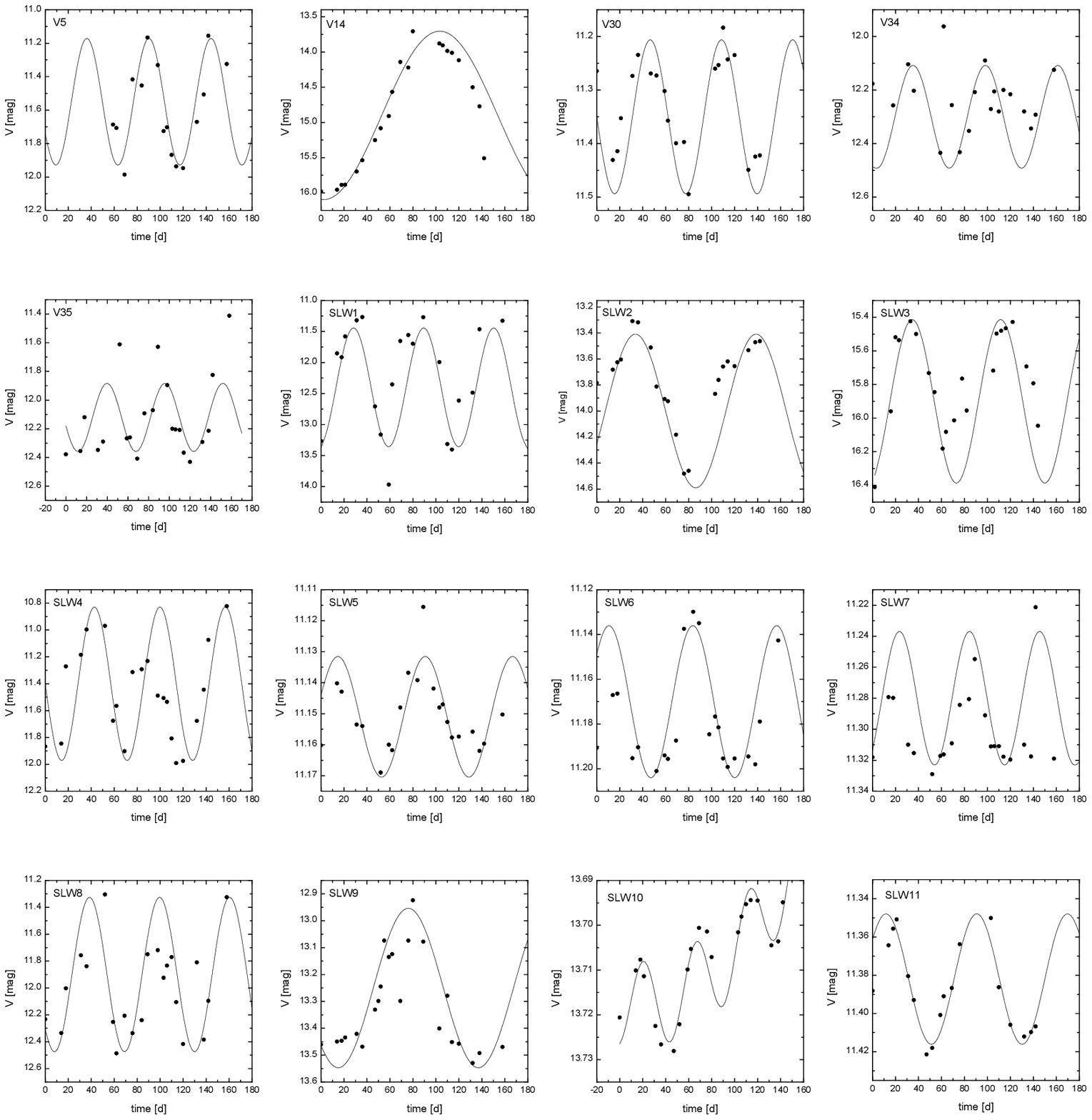}
\caption{Light change of the long period variables in M22 plotted against time. Starting date is JD 2452420.}
\label{M22plot}
\end{center}
\end{figure*}

The light variations of the LPVs in M22 are presented in Figure \ref{M22plot} plotted against time starting with the
first date of our time series. They are compared with synthetic light curves based on the period given in Table\,\ref{tab1}.
Photometry data are given in Table \ref{tab4}.
SLW10 shows a very long period variation in addition to the 46~d period. 
Such long periods are a well known phenomenon in red variables \cite{P03,K99}, but its nature is not understood yet \cite{W04}.
For illustrative purposes, a long time trend has been added also to the
synthetic light curve.
SLW10 is also interesting since it is possibly identical 
with an X-ray source. 

It is quite 
obvious that most stars show some kind of irregular behaviour on top of the adopted periodic light change
as expected for semiregular variables. We note that the period of SLW9, although it seems to give a nice fit 
is classified as uncertain since we could not detect a second maximum.
The light curve of V34 is very unusual for an LPV. Plotted against phase it rather resembles the one expected for
an eclipsing binary. We therefore suspect that for this star the observed variability is caused by a companion.

In addition to the variables 
for which we could estimate a period, we list in Table\,\ref{tab2} 
several red stars that show variability but without a clear periodicity. This may be due to a truly irregular
nature of their light change or a larger photometric error, e.g. if there is another star nearby.
We did not investigate these cases further.

A proper motion study for individual stars in NGC 6656 can be found in the paper by Zloczewski et 
al.~\shortcite{Zl12}. According to these authors, V9, V34, and SLW6 are likely cluster members,
V30, SLW11, and SLW20 are probably non-members. However, V30 is listed as a member based on its radial
velocity measured by Peterson and Cudworth \shortcite{PC94}. These authors find a
mean radial velocity of approximately 147.6$\pm$9.8\,km/s for a sample of 
170 objects they consider 
cluster members. Their list gives also support to the cluster membership of V5, V9, and SLW4.
We note that the LPV candidate V8 listed by Clement et al. \shortcite{Cl01}, which was not observed within our program,
fulfills both membership criteria as well. 

\begin{figure}
\begin{center}
\includegraphics[width=19pc,height=19pc]{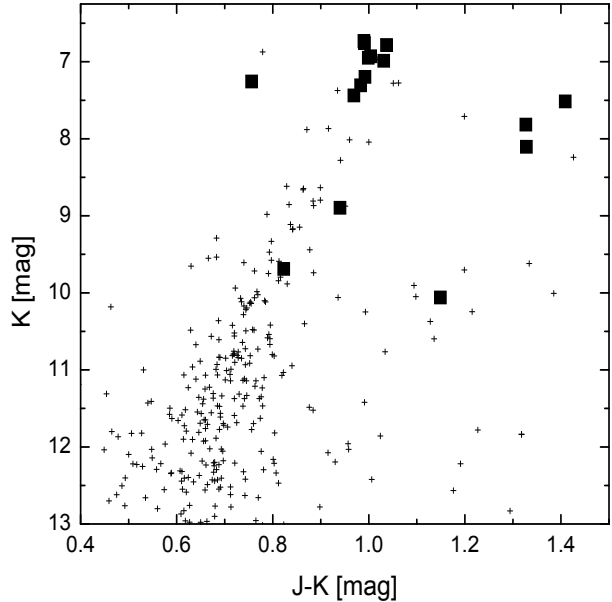}
\caption{Colour-magnitude diagram of the central part of M22 using 2MASS $J$ and $K$ photometry. The filled
boxes mark the variables from Table \ref{tab1}.}
\label{CMDM22}
\end{center}
\end{figure}

From the 2MASS $J$ and $K$ photometry, we find that V5, V9, V30, V34, V35, SLW1, SLW4, SLW5, SLW7, SLW8, SLW9,
and SLW11 are located on the cluster's giant branch (see Fig.\,\ref{CMDM22}). 
SLW1 and SLW9 are more than 2 magnitudes below the tip of
the giant branch and are thus likely RGB stars. It is unlikely that stars with their relatively large
amplitudes would be found this far below the RGB tip, so they are likely to be non-members. 
SLW6 is offset to the blue in $(J-K)$, but its $(V-I)$ colour is similar to
the other red giants.
Since it is probably
a cluster member due to its proper motion \cite{Zl12}, we suspect that the 2MASS $K$ brightness of this star may be incorrect.
V14, SLW2, SLW3, and SLW10 are found on the right side of the giant branch. It is difficult to decide on
their cluster membership since their colour and brightness may be affected by circumstellar dust or their
variability (in particular for the large amplitude variable V14). However, they may also be background sources.

\subsection{IC4499}
\begin{figure}
\begin{center}
\includegraphics[width=19pc,height=19pc]{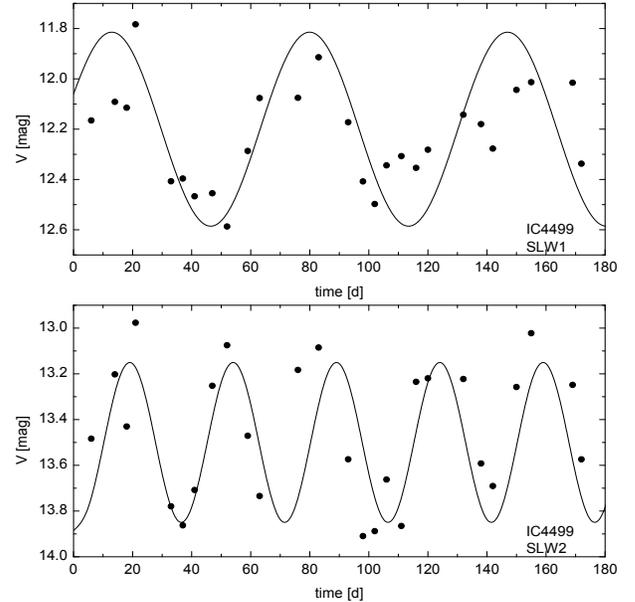}
\caption{Light change of the two newly detected variables in IC 4499 plotted against time. Starting date is JD 2452420.}
\label{ic4499lc}
\end{center}
\end{figure}

In the MACHO data of the second cluster, 
two objects were found to have significant variability, both previously unreported. They are 
listed in Table \ref{tab3} and their light curves plotted against time are presented in Figure \ref{ic4499lc}.
Photometry data for both stars can be found in Table \ref{tab5}.
No long secondary period has been found in any one of them, both of them are monoperiodic.

No proper motion data have been reported in literature for IC4499. 
Hankey and Cole \shortcite{HC11} published a radial velocity study of stars in that field, but none of our
LPV candidates was observed. From their $K$ brightness, both variables are found
at the top of the red giant branch, and they are within a 5 arcmin radius of the cluster centre. 
However, their $J-K$ colour is slightly too blue compared with the
bulk of probable RGB stars in IC4499 \cite{HC11}. SLW1 is also too blue in $V-I$ for an RGB star according
to the photometry presented by Walker and Nemec \shortcite{WN96}, while SLW2 has an appropriate $V-I$
value. For both stars, their brightness in $V$ places them above the cluster's giant branch.
Their membership is therefore not clear, and a possible
coincidence of their location close to the cluster centre has to be counterbalanced with their too blue colour.
As Hankey and Cole noted, the cluster is seen through the outer parts of the Galactic bulge, therefore it cannot
be excluded that these objects are actually bulge stars. Due to their brightness and colour we tend to classify
both objects as non-members.


\section{DISCUSSION}
For both clusters studied here,
the fields observed will likely contain a considerable contribution from field (or bulge) stars. To discuss
the long period variables in M22 and IC4499 we thus have to distinguish cluster members and field stars first.
As shown above, velocity and proper motion data found in the literature do not allow a clear separation in
all cases. We decided to do a first selection of members on the basis of their location in the 2MASS 
colour-magnitude diagrams (Fig.\,\ref{CMDM22} for M22; see Hankey and Cole \shortcite{HC11} for IC4499).
As a next step, we did a cross-check of these candidates using their pulsational properties.

\begin{figure}
\begin{center}
\includegraphics[width=19pc,height=19pc]{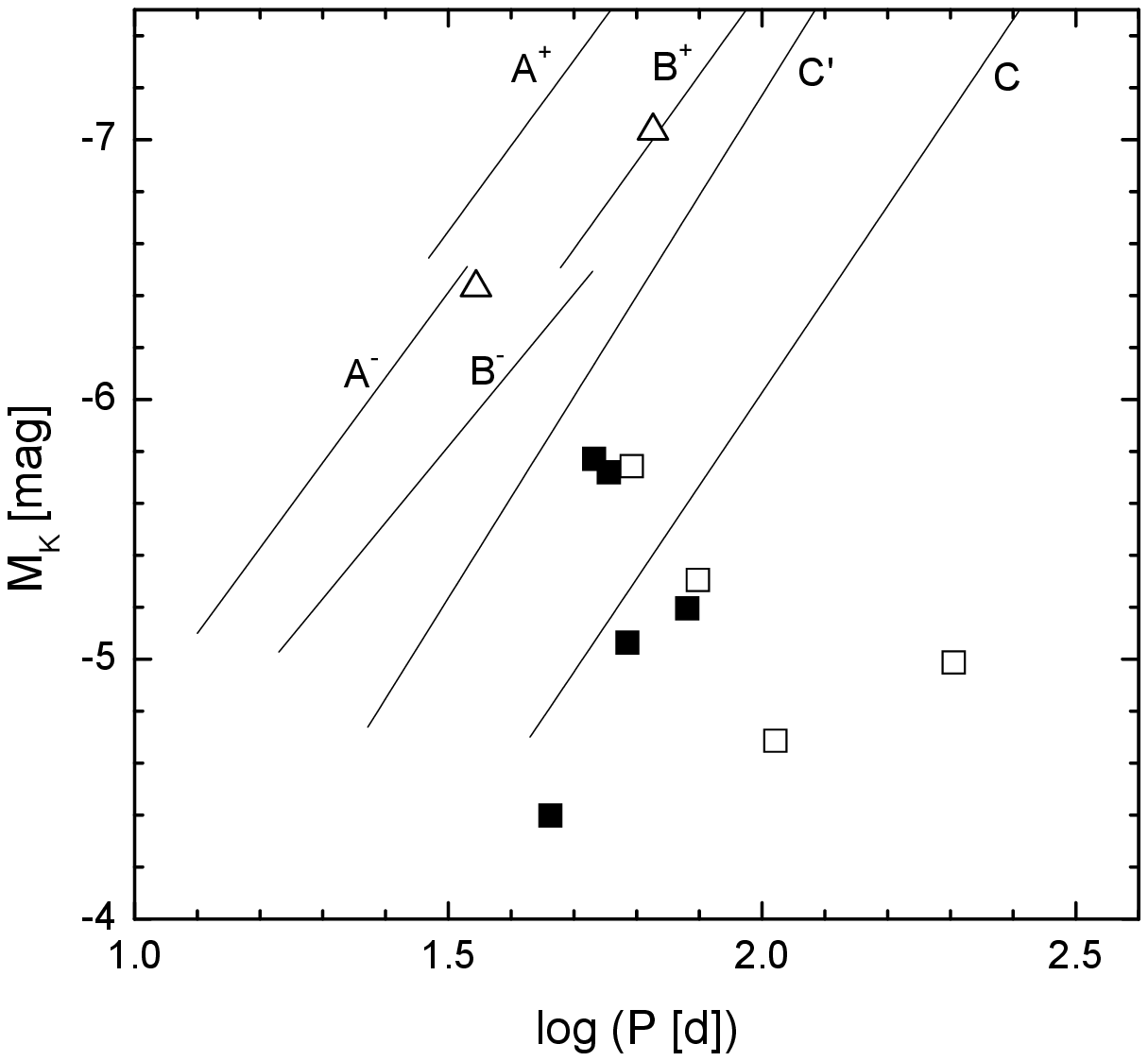}
\caption{LogP-$K$-diagram of the variables found in this study. Boxes denote stars in M22 while the two objects
in IC4499 are marked with triangles. Open symbols indicate uncertain membership. Only stars with reliable
periods based on our data are plotted. Furthermore, the plot gives the location of the logP-$K$-relations for
LPVs derived from the Large Magellanic Cloud by Ita et al.\,\shortcite{ita04}.}
\label{logPK}
\end{center}
\end{figure}

In Fig.\,\ref{logPK} we show a logP-$M_{K}$-diagram for the variables listed in Tabs.\,\ref{tab1} and
\ref{tab3} with reliable periods. 
Absolute magnitudes have been calculated using a distance modulus of 12.46 and 16.23 for M22
and IC4499, respectively. Reddening values E(B-V) of 0.38 and 0.23 have been used. Interstellar extinction
in $K$ has been calculated using the relation given in Cardelli et al. \shortcite{CCM89}. We also
added the logP-$K$-relations for the LMC given by Ita et al. \shortcite{ita04} shifted with a distance
modulus of 18.5 and transformed from the LCO to the 2MASS system following Carpenter \shortcite{carp01}.
Variables with uncertain cluster membership are marked by open symbols.

Based on the logP-$K$-diagram shown in Fig.\,\ref{logPK}
we further exclude SLW1\footnote{Not visible in Fig.\,\ref{logPK} due to its low $K$ brightness},
SLW2 and the large amplitude variable V14. Since the period is also used as a
selection criterion now, we remove all variables with unknown or uncertain period as well. 
We note that the period from Wehlau and Sawyer-Hogg \shortcite{WS77}
would place the star V9 (Table \ref{tab2}) close to the other M22 variables in this diagram.
Our resulting sample includes 7 stars, namely V5, V30, SLW4, SLW5, SLW7, SLW10, and SLW11.
The membership of SLW11 and V30 is somewhat uncertain according to the literature data, but we decided to
leave them in the sample, too. From their brightness, the stars are either on
the RGB or on the AGB. 
In our second cluster, IC4499, the membership of both LPV candidates is doubtful due
to their brightness and colour. 

Among the clusters studied so far within our monitoring program, M22 and IC4499 are the two with the lowest 
metallicity. Nevertheless, LPVs are found at least in one of them, M22. Among our 7 good 
candidates, 4 have $V$ amplitudes of a few tenth of a magnitude or more. We remind the reader that a mira-like
variability with a $V$ amplitude of several magnitudes is directly related to the high temperature sensitivity of
the TiO bands in the visual part of the spectrum and therefore not expected for low metallicity stars like the
ones we study here. This is in agreement with the typical amplitudes found in LPVs in the somewhat more metal
rich clusters NGC 362 and NGC 2808 \cite{LW11}. Considering the detection of LPVs in the even more metal poor
cluster M15 by McDonald et al. \shortcite{McD10} we propose that this kind of variability appears at the upper
end of the giant branch also at much lower metallicity than expected previously \cite{FE88}.

Beside TiO playing less of a role for the visual amplitude, in low metallicity clusters the long period variables
are found at a lower luminosity compared to, e.g., the LPVs in 47 Tuc. This trend seen already in NGC 362 and 
NGC 2808 is clearly continued in M22 with the most luminous cluster LPV having $M_{K}$ below $-$6, more than
one magnitude below the most luminous LPVs in 47 Tuc. At the same time, the LPV phenomenon sets in at a lower
brightness than in 47 Tuc. Should the two variables found in the field of IC4499 be indeed cluster members, this
would naturally contradict the findings from the other clusters. 
It could be explained if this cluster is indeed younger than clusters like M22, but this option has been
rejected in the most recent study of this issue by Walker et al.~\shortcite{WKA11}.
Therefore, it would be important to clarify their
cluster membership or non-membership unambiguously. That we did not find any LPV candidates in IC4499 at a
lower luminosity is probably due to the cluster's distance and significantly smaller size in terms of member
stars. 

Finally, we take a look at the location of the 7 M22 LPVs relative to the LPV logP-$K$-relations found for the LMC
(Fig.\,\ref{logPK}). We find four stars close to sequence C, which has been related to fundamental mode pulsation,
and three stars between sequence C and C'. 
The LPV candidate V8 from Clement et al. \shortcite{Cl01}
would be placed close to sequence C as well using its literature period (61\,d) and 2MASS $K$-magnitude.
However, that star is slightly shifted to the blue in $J-K$ relative to the cluster's giant branch, so that its classification
as LPV is not totally clear. 

Our variables seem to be somewhat offset from the sequences -- shifted
either to a longer period or a lower brightness. From our previous work \cite[in particular]{LW11} we know
that both low metallicity and mass loss tend to shift the logP-$K$-relations to the right in such diagrams, in
agreement with the very low metallicity of the M22 stars. The alternative explanation would be a small uncertainty
in the distance to M22, which seems more unlikely. An interesting feature of the diagram is that the most
luminous LPVs in M22 are not found on sequence C but on sequence C'. The low number of variables with reliable data
prohibits the drawing of further conclusions on this finding. 

To sum up, we report the detection of 24 previously unknown 
long period variables in the fields of the two metal poor 
globular clusters M22 and IC 4499. Periods were derived for 17 of the LPVs
and cluster membership was discussed. The detection of
LPVs as members of these metal poor clusters adds to the sketching of a complete picture of variability on the
upper giant branch at various metallicities.

\begin{table*}
\caption{Visual magnitudes of the long period variables in the field of NGC 6656.}
\begin{center}
\begin{tabular*}{\textwidth}{@{}l\x c\x c\x c\x c\x c\x c\x c\x  c@{}}
\hline\hline
Julian Date & V5 & V14 & V30 & V34 & V35 & SLW1 & SLW2 & SLW3 \\
~ & [mag] & [mag] & [mag] & [mag] & [mag] & [mag] & [mag] & [mag] \\
\hline%
2452420 & 11.959 & 15.980 & 11.265 & 12.177 & 12.378 & 13.270 & 13.784 & 16.410 \\
2452434 & 12.000 & 15.957 & 11.431 & 12.731 & 12.356 & 11.853 & 13.683 & 15.960 \\
2452438 & 11.439 & 15.887 & 11.414 & 12.258 & 12.120 & 11.915 & 13.625 & 15.519 \\
2452451 & 11.709 & 15.886 & 11.353 & 12.104 & 12.349 & 11.580 & 13.604 & 15.537 \\
2452456 & 11.825 & 15.698 & 11.274 & 12.203 & 12.289 & 11.320 & 13.309 & 15.424 \\
2452472 & 11.054 & 15.538 & 11.235 & 11.196 & 11.612 & 11.268 & 13.319 & 15.499 \\
2452479 & 11.687 & 15.253 & 11.269 & 12.435 & 12.267 & 12.708 & 13.512 & 15.732 \\
2452482 & 11.707 & 15.084 & 11.273 & 11.963 & 12.261 & 13.162 & 13.812 & 15.846 \\
2452489 & 11.986 & 14.911 & 11.302 & 12.257 & 12.409 & 13.969 & 13.908 & 16.182 \\
2452496 & 11.417 & 14.567 & 11.357 & 12.433 & 12.093 & 12.354 & 13.925 & 16.082 \\
2452504 & 11.453 & 14.142 & 11.399 & 12.353 & 12.071 & 11.653 & 14.182 & 16.014 \\
2452509 & 11.166 & 14.221 & 11.397 & 12.208 & 11.629 & 11.559 & 14.482 & 15.765 \\
2452518 & 11.331 & 13.708 & 11.495 & 12.090 & 11.896 & 11.697 & 14.459 & 15.955 \\
2452523 & 11.726 & 13.880 & 11.260 & 12.271 & 12.200 & 11.271 & 13.868 & 15.717 \\
2452526 & 11.703 & 13.906 & 11.253 & 12.206 & 12.205 & 11.993 & 13.761 & 15.497 \\
2452530 & 11.868 & 13.985 & 11.184 & 12.280 & 12.209 & 13.314 & 13.659 & 15.480 \\
2452534 & 11.936 & 14.012 & 11.243 & 12.200 & 12.367 & 13.405 & 13.619 & 15.465 \\
2452540 & 11.947 & 14.117 & 11.235 & 12.216 & 12.432 & 12.616 & 13.655 & 15.428 \\
2452552 & 11.671 & 14.503 & 11.449 & 12.280 & 12.293 & 12.487 & 13.533 & 15.693 \\
2452558 & 11.507 & 14.773 & 11.425 & 12.344 & 12.214 & 11.465 & 13.471 & 15.794 \\
2452562 & 11.156 & 15.509 & 11.422 & 12.292 & 11.825 & 11.402 & 13.463 & 16.046 \\
2452578 & 11.325 & 15.537 & 11.398 & 12.125 & 11.412 & 11.326 & 13.492 & 16.326 \\
\hline \hline \\
\hline\hline
Julian Date & SLW4 & SLW5 & SLW6 & SLW7 & SLW8 & SLW9 & SLW10 & SLW11 \\
~ & [mag] & [mag] & [mag] & [mag] & [mag] & [mag] & [mag] & [mag] \\
\hline%
2452420 & 11.867 & 11.160 & 11.191 & 11.318 & 12.234 & 13.461 & 13.721 & 11.388 \\
2452434 & 11.846 & 11.140 & 11.167 & 11.279 & 12.336 & 13.449 & 13.710 & 11.364 \\
2452438 & 11.271 & 11.143 & 11.166 & 11.280 & 12.003 & 13.447 & 13.708 & 11.356 \\
2452441 & 11.185 & 11.153 & 11.195 & 11.310 & 11.757 & 13.435 & 13.711 & 11.351 \\
2452451 & 10.997 & 11.154 & 11.190 & 11.315 & 11.839 & 13.421 & 13.722 & 11.380 \\
2452456 & 10.970 & 11.169 & 11.201 & 11.329 & 11.305 & 13.468 & 13.727 & 11.393 \\
2452467 & 11.676 & 11.160 & 11.194 & 11.317 & 12.254 & 13.331 & 13.728 & 11.421 \\
2452472 & 11.565 & 11.162 & 11.196 & 11.316 & 12.487 & 13.245 & 13.722 & 11.418 \\
2452479 & 11.902 & 11.148 & 11.187 & 11.309 & 12.207 & 13.135 & 13.710 & 11.401 \\
2452482 & 11.313 & 11.137 & 11.138 & 11.284 & 12.337 & 13.124 & 13.705 & 11.391 \\
2452489 & 11.293 & 11.139 & 11.130 & 11.281 & 12.240 & 13.299 & 13.701 & 11.387 \\
2452496 & 11.231 & 11.116 & 11.135 & 11.259 & 11.749 & 13.074 & 13.701 & 11.364 \\
2452500 & 11.488 & 11.142 & 11.185 & 11.291 & 11.719 & 12.924 & 13.707 & 11.323 \\
2452523 & 11.506 & 11.148 & 11.177 & 11.311 & 11.925 & 13.078 & 13.702 & 11.350 \\
2452526 & 11.535 & 11.147 & 11.182 & 11.311 & 11.833 & 13.401 & 13.698 & ~      \\
2452530 & 11.807 & 11.153 & 11.195 & 11.311 & 11.770 & 13.279 & 13.695 & 11.386 \\
2452534 & 11.991 & 11.158 & 11.199 & 11.318 & 12.106 & 13.451 & 13.694 & ~     \\
2452540 & 11.975 & 11.157 & 11.195 & 11.320 & 12.418 & 13.458 & 13.694 & 11.406 \\
2452552 & 11.677 & 11.156 & 11.194 & 11.310 & 11.810 & 13.529 & 13.704 & 11.412 \\
2452558 & 11.445 & 11.162 & 11.198 & 11.318 & 12.385 & 13.493 & 13.704 & 11.410 \\
2452562 & 11.074 & 11.160 & 11.179 & 11.221 & 12.096 & 13.481 & 13.695 & 11.407 \\
2452578 & 10.823 & 11.150 & 11.143 & 11.319 & 11.325 & 13.470 & 13.690 & ~ \\
\hline \hline
\end{tabular*} \label{tab4}
\end{center}
\end{table*}

\begin{table}
\caption{Visual magnitudes of the long period variables of IC 4499.} 
\begin{center}
\begin{tabular}{lcc}
\hline \hline
Julian Date & SLW1 & SLW2 \\
~ & [mag] & [mag] \\
\hline 
2452426 & 12.165 & 13.484  \\
2452434 & 12.091 & 13.202 \\
2452438 & 12.115 & 13.430 \\
2452441 & 11.783 & 12.976 \\
2452453 & 12.407 & 13.779 \\
2452457 & 12.396 & 13.863 \\
2452461 & 12.467 & 13.708 \\
2452467 & 12.455 & 13.252 \\
2452472 & 12.587 & 13.075 \\
2452479 & 12.287 & 13.471 \\ 
2452483 & 12.076 & 13.734 \\
2452496 & 12.075 & 13.183 \\
2452503 & 11.914 & 13.085 \\
2452513 & 12.172 & 13.574 \\
2452518 & 12.408 & 13.909 \\
2452522 & 12.498 & 13.888 \\
2452526 & 12.344 & 13.662 \\
2452531 & 12.307 & 13.865 \\
2452536 & 12.354 & 13.235 \\ 
2452540 & 12.282 & 13.220 \\
2452552 & 12.143 & 13.223 \\
2452558 & 12.180 & 13.592 \\
2452562 & 12.277 & 13.690 \\
2452570 & 12.044 & 13.258 \\
2452575 & 12.013 & 13.022 \\
2452589 & 12.015 & 13.242 \\
2452592 & 12.337 & 13.574 \\
\hline \hline
\end{tabular} \label{tab5}
\end{center}
\end{table}

\begin{acknowledgements}
TL acknowledges support by the Austrian Science Fund under project number P23737-N16.
The authors wish to thank Warren Hankey for providing radial velocity data for IC4499.
\end{acknowledgements}

\end{document}